\documentclass[aps,prd,preprintnumbers,nofootinbib,amsmath,amssymb]{revtex4}

\begin{document}

\preprint{CCNY-HEP-03/7}

\title{Hamiltonian analysis of the noncommutative Chern-Simons theory}

\date{\today}

\author{Alexandr Yelnikov}
\altaffiliation[Also at ]{Department of Physics, City College of the CUNY, New York, NY 10031}
\affiliation{Department of Physics, Queens College of the CUNY, Flushing, NY 11367}
\email{yelnikov@qc.edu}

\begin{abstract}
In this paper the hamiltonian analysis of the pure Chern-Simons theory on the noncommutative plane is performed. We use
the techniques of geometric quantization to show that the classical reduced phase space of the theory has nontrivial
topology and that quantization of the symplectic structure on this space is possible only if the Chern-Simons coefficient is
quantized.
Also we show that the physical Hilbert space of the theory is one dimensional and give an explicit expression for the
vacuum wavefunction. This vacuum state is found to be related to the noncommutative Wess-Zumino-Witten action.

\end{abstract}

\maketitle
\section{Introduction}
Chern-Simons (CS) theories have been extensively investigated in
various contexts since their appearance in physics literature as
topological mass terms for odd dimensional gauge theories
\cite{Deser}.  On the other hand, recent progress in understanding
connection between string theory and noncommutative geometry
\cite{Connes} has brought much attention to studies of the field
theories over the noncommutative spaces. In particular,
noncommutative version of the Chern-Simons  theory have been
proposed in both star-product \cite{Cham} and operator formalism
\cite{Poly}.  And although this theory has been discussed by many
authors by now \cite{Chen}, much of the previous analysis was done
using conventional Lagrangian formalism and path-integral
quantization techniques. It is, however, well-known that at least
in the commutative case, hamiltonian approach has been much more
useful in illuminating various aspects of the CS theory. Therefore
it appears to be interesting to extend the canonical formalism to
the noncommutative Chern-Simons theory (NCCS) as well.

Also as it was found recently in \cite{Susskind}, the
noncommutative $U(1)$ Chern-Simons theory  can be quite useful in
describing fractional quantum Hall effect.  The argument of that
paper crucially depends on whether Chern-Simons coefficient (also
known as level number) is quantized or not. After some initial
controversy \cite{Sheikh} it was finally shown in \cite{Nair} that
noncommutative CS theory shares the same property as its
commutative counterpart, {\textit {i.e.}} the quantization of the
level number. In fact the result is even stronger in the
noncommutative case. CS coefficient is quantized even for the
$U(1)$ theory indicating that as the noncommutativity parameter
$\theta$ approaches zero, the noncommutative $U(1)$ Chern-Simons
theory does not go over smoothly to the commutative one.

In the Lagrangian formalism the reason for level quantization is standard. Like in the commutative case, one can show
{\cite{Nair}} that NCCS action is not invariant under the gauge transformations belonging to nontrivial homotopy classes of the
gauge group ${\cal G}$. For transformation with winding number $n$ the action changes by $8\pi^2\lambda n$ and the
requirement of single-valuedness of the path-integral measure leads to the following quantization condition on the Chern-Simons
coefficient $\lambda$
\begin{equation}\label{qcondition}
\lambda = \frac{n}{4\pi}, \ \ \ \ \ n=\pm 1, \pm 2 \ldots.
\end{equation}

In this paper we would like to consider canonical quantization of
the pure $U(N)$ Chern-Simons theory on the noncommutative plane
and to show how this quantization condition appears in the
Hamiltonian formalism. In Section $2$ we give the classical
analysis of the phase space of the theory in the framework of
geometric quantization. In particular, we show that the reduced
phase space is topologically nontrivial and therefore consistent
quantization of the symplectic structure on this space is possible
only if the level number is quantized. Section $3$ describes
canonical quantization of the theory in the functional
Schr\"odinger representation. We find that consistent realization
of  the Gauss law constraint on the wave functionals leads to
nontrivial transformation law of physical states under the gauge
transformations. And like in the commutative case, this
transformation law combined with single-valuedness of wave
functionals gives us once again the Chern-Simons coefficient
quantization condition (\ref{qcondition}). In Section $4$ we
explain how to construct the most general functionals obeying the
gauge transformation law mentioned above. It turns out that the
physical Hilbert space for our choice of the flat space geometry
is one-dimensional. We find an explicit expression for the only
physical state of the theory and show that it is naturally
connected to the noncommutative Wess-Zumino-Witten action. At the
end of this section we  describe how the results of the ordinary
commutative CS theory can be recovered from our expressions in the
limit of vanishing noncommutativity parameter $\theta$.

Throughout this paper we use the following conventions. The
noncommutative plane is defined in the usual way~\cite{Douglas} by
introducing the coordinate operators $(x_1, x_2)$ which satisfy the
commutation relation
\begin{equation}
[x_i, x_j] = i\theta \delta_{ij} \ \ \ \ \ i,j =1,2,
\end{equation}
where $\theta$ is the {\textit c}-number parameter characterizing
the noncommutativity of space. The actual space can be thought
of as a representation of the operator algebra generated by $(x_1, x_2)$,
and the standard realization of the noncommutative plane is given
by the Fock oscillator basis $|n\rangle, n=0, 1,2,\ldots$
\begin{eqnarray}
\bar z|n\rangle & = & \sqrt{2\theta}\,\sqrt{n}\,
|n+1\rangle\\\nonumber z|n\rangle & =
&\sqrt{2\theta}\,\sqrt{n-1}\, |n-1\rangle\\\nonumber z|0\rangle &
=& 0\\\label{ccoord}
z = x_1 + i x_2 & & \bar z = x_1 - i x_2.
\end{eqnarray}
Functions on this space are the elements of the enveloping algebra of $(x_1, x_2)$,
while  derivatives are given  by the inner automorphisms of that
algebra
\begin{equation}\label{ncder}
\partial_i(\ldots) = [\hat\partial_i, \ldots].
\end{equation}
These automorphisms are generated by derivative operators
\begin{equation}
\hat\partial_i = \frac{i}{\theta}\epsilon_{ij} \hat x_j
\end{equation}
or
\begin{equation}
\hat \partial = -\frac{\bar z}{2\theta} \ \ \ \ \ \hat\bar\partial = \frac{z}{2\theta}
\end{equation}
if we use complex coordinates (\ref{ccoord}). Since we are going to analyze
noncommutative Chern-Simons theory in operator formulation, it is
convenient to think of the functions as infinite matrices with raw
and column indices labeling the oscillator states. In the case of
the $U(N)$ theory (as opposed to $U(1)$) we should take the direct
sum of $N$ copies of the Fock space (which is certainly isomorphic
to a single space) and consider all the relevant quantities (gauge
connections, covariant derivative operators, {\it etc}) as
operator-valued $N\times N$ matrices or as infinite matrices with
double index labeling both oscillator states and internal degrees
of freedom. The space integral on the commutative plane becomes a trace in
the noncommutative case
\begin{equation}
\int d^2 x \ \longrightarrow \ 2\pi\theta\ {\rm Tr}
\end{equation}
and for $U(N)$ theory ${\rm Tr}$ stands for the integration over the noncommutative plane as well as for the $U(N)$ group trace.

\section{Canonical formalism}
The starting point of our analysis is the matrix form of the noncommutative Chern-Simons action  proposed in \cite{Poly}
\begin{equation}\label{NCCS}
S_{NCCS} = 2\pi i\theta\lambda \int dt {\rm Tr}(\frac{2}{3} D_{\mu}D_{\nu}D_{\kappa} \epsilon^{\mu\nu\kappa}) +
4\pi\lambda \int dt {\rm Tr} D_0 .
\end{equation}
Here $D_\mu, \mu =0,1,2$ - are hermitian matrix-valued covariant
derivative operators, transforming adjointly under the $U(N)$
gauge transformations
\begin{equation}\label{gauge}
D_\mu \longrightarrow D_\mu^U = U D_\mu U^{-1}.
\end{equation}
These operators are related to ordinary noncommutative gauge connections $A_\mu, \mu =0,1,2$ {\textit via}
\begin{eqnarray}\label{DArelation}
D_i & = & -i \hat \partial + A_i \ \ \ \ \ \ i=1,2\\ \nonumber
D_0 & = & -i\partial_t + A_0.
\end{eqnarray}
For canonical quantization purposes it is most convenient to choose the time-axial gauge $A_0=0$. In this gauge, the action (\ref{NCCS})
is quadratic in $D_1, D_2$
\begin{equation}
S_{NCCS} = 2\pi\theta\lambda \int dt {\rm Tr}(\partial_t D_1 D_2 - \partial_t D_2 D_1)
\end{equation}
and since it is first order in time derivatives, we can
immediately write the symplectic $2$-form $\Omega$
\begin{equation}
\Omega = 8\pi\theta\lambda i \ {\rm Tr}(\delta Z \delta \bar Z)\\ \label{ccr}
\end{equation}
as well as Poisson brackets
\begin{equation}
\{Z_{ij}, \bar Z_{kl}\} = \frac{i}{8\pi\theta\lambda}\,\delta_{il}\,\delta_{jk}\label{poisson}
\end{equation}
on the space of all covariant derivatives~$\cal Z$. Here we
introduced the complex coordinates $Z = \frac{1}{2}(D_1 -i D_2),\
\bar Z = \frac{1}{2}(D_1 +i D_2)$ on $\cal Z$, and $\delta$ in
(\ref{ccr}) is to be interpreted as denoting exterior
derivative on this space (we do not write the wedge sign for
exterior products on $\cal Z$ since it is clear from the context).
With this choice of coordinates $\cal Z$ can be considered as a
K\"ahler manifold with $\Omega$ being the K\"ahler form and
\begin{equation}\label{Kahler}
K= 8\pi\theta\lambda i {\rm Tr}(Z \bar Z)
\end{equation}
being the K\"ahler potential.

Moreover, the Hamiltonian is obviously zero meaning that there is no time evolution in this theory. Equivalently, we can say
that equations of motion for $Z, \bar Z$
\begin{equation}
\partial_t Z = 0  \ \ \ \ \ \ \partial_t \bar Z =0
\end{equation}
are satisfied trivially with time-independent matrices. However, these matrices have
to satisfy an extra constraint (the Gauss law constraint)
\begin{equation}\label{Aeq}
\theta [Z, \bar Z] +\frac{1}{2} =0
\end{equation}
which appears from (\ref{NCCS}) as an equation of motion for $A_0$. Equation (\ref{Aeq}) is a matrix identity, although matrix
indices are suppressed for simplicity. It is also easy to see that this  equation can not be satisfied with finite
matrices, meaning that we consider an essentially infinite-dimensional matrix model. 

At this point we have the two alternatives. One may impose the
Gauss law constraint at the classical level. This yields the
reduced phase space, the set of matrices $Z, \bar Z$ satisfying
(\ref{Aeq}) up to gauge transformations, endowed with a symplectic
structure inherited from (\ref{ccr}). This reduced phase space may
be quantized using the holomorphic polarization induced by complex
structure on the noncommutative plane. In this paper, however, we
will follow an alternative procedure of quantizing the Poisson
brackets (\ref{poisson}) first. In this case $\cal Z$ may be
considered as the phase space of the theory before reduction by
the action of the gauge symmetries. Reduction is done by requiring
that Gauss law constraint acts on the Hilbert space thus selecting
the subspace of physical states. As will be shown shortly, the
quantization of the level number appears in this case as a
consistency condition for performing such reduction.

But before we can proceed a few words about gauge transformations
are in order. The $A_0 =0$ condition does not fix the gauge
completely: one can still make time-independent gauge
transformations. However, we have to be careful about what the
allowed gauge transformations are. We want to show now that the
group $\cal G$ of the valid gauge transformations is given by
those unitary matrices only, which act as identity on the
oscillator basis $|n\rangle$ as $n\to\infty$. This property is the
noncommutative version of the requirement that gauge
transformations go to identity at spatial infinity.

For infinitesimal gauge transformation $U = 1+\phi +\ldots$
($\phi$ - antihermitian matrix) we obtain from (\ref{gauge})
\begin{eqnarray}\label{infgauge}
\delta Z & = &[\phi, Z]\\ \nonumber
\delta \bar Z & = & [\phi, \bar Z].
\end{eqnarray}
The vector field on ${\cal Z}$ generating such transformation is
\begin{equation}
\xi = [\phi, Z]_{ij} \frac{\delta}{\delta Z_{ij}} + [\phi, \bar Z]_{ij} \frac{\delta}{\delta \bar Z_{ij}}.
\end{equation}
By contracting this with $\Omega$ we get
\begin{eqnarray}
i_{\xi}\Omega & = & 8\pi\theta\lambda i {\rm Tr}([\phi, Z] \delta \bar Z -[\phi, \bar Z] \delta Z)\\ \nonumber
& = & 8\pi\theta\lambda i {\rm Tr} (\phi [Z, \delta \bar Z] + \phi [\delta Z, \bar Z])\\ \nonumber
& = & 8\pi\theta\lambda i {\rm Tr} (\phi\, \delta [Z, \bar Z]).
\end{eqnarray}
This identifies the generator $G$ of  infinitesimal gauge transformation (\ref{infgauge}) up to an arbitrary constant as
\begin{equation}
G_{ij} = 8\pi\theta\lambda i [Z, \bar Z]_{ij} + {\rm const}\, \delta_{ij}.
\end{equation}
We can fix this constant by requiring that $G(\phi)=0$ condition is equivalent to the Gauss law constraint (\ref{Aeq}). 
Therefore, the fixed expression for $G(\phi)$ is
\begin{equation}\label{gener}
G(\phi) = 8\pi\lambda i {\rm Tr}\, \phi(\theta [Z, \bar Z] + \frac{1}{2}).
\end{equation}
As a consistency check we can evaluate, using the canonical Poisson brackets (\ref{poisson}), the commutator of two such
 transformations with infinitesimal parameters $\phi$ and $\rho$
\begin{equation}
[G(\phi), G(\rho)] = -i G([\phi, \rho]) -4\pi\lambda {\rm Tr}[\phi, \rho].
\end{equation}
From this expression we see that the algebra of these generators gives the representation of the Lie algebra of the gauge group ${\cal G}$ provided
\begin{equation}\label{gaugecondition}
{\rm Tr} [\phi, \rho] = 0.
\end{equation}
This last condition is satisfied only by functions $\phi, \rho$ which act as zero on the oscillator basis states $|n\rangle$ for large $n$. In this case,
$\phi$ and $\rho$ are essentially finite matrices and we can use the cyclicity of trace to prove (\ref{gaugecondition}). This
also validates the statement we have made above that closure of the algebra of gauge transformations restricts ${\cal G}$ only
to those unitary matrices which go to identity at spatial infinity.

Once we have identified the group ${\cal G}$  we would like to
analyze the reduced phase space ${\cal Z}/{\cal G}$ of covariant
derivatives modulo gauge transformations in more detail. We want
to show now that this space has nontrivial topology, in
particular, that there are closed noncontractible two-surfaces in
${\cal Z}/{\cal G}$.

Let $(Z_0, \bar Z_0)$ denote a specific set of matrices
corresponding to a point in ${\cal Z}$. Consider now the
$2$-surface in this space parameterized by $\sigma$ and $\tau$
\begin{eqnarray}
Z = (1-\sigma) Z_0 + \sigma U Z_0 U^{-1}\\ \nonumber
\bar Z = (1-\sigma) \bar Z_0 + \sigma U \bar Z_0 U^{-1}
\end{eqnarray}
where $0\leq \sigma \leq 1$ and $U(\tau)$ is a one-parameter family of gauge transformations with $U(0)=U(1)\equiv {\bf 1}$,
 $0\leq \tau \leq 1$. Easy to see that in the reduced phase space ${\cal Z}/{\cal G}$, the boundary of this surface corresponds to the single
point $(Z_0, \bar Z_0)$ and so we have a closed $2$-surface in ${\cal Z}/{\cal G}$. This closed surface is not contractible
if $U(\tau)$ traces a
noncontractible path in ${\cal G}$. We can now integrate the symplectic $2$-form $\Omega$ over this surface:
\begin{eqnarray}
\delta Z = \delta \sigma (U Z_0 U^{-1}- Z_0) - \sigma U [Z_0, U^{-1}\delta U] U^{-1}\\ \nonumber
\delta \bar Z = \delta \sigma (U \bar Z_0 U^{-1}- \bar Z_0) - \sigma U [\bar Z_0, U^{-1}\delta U] U^{-1}
\end{eqnarray}
so
\begin{eqnarray}
\int \Omega &=& 8\pi\theta\lambda i \int \delta\sigma\, \sigma {\rm Tr}\left[(\bar Z_0 -U^{-1} \bar Z_0 U)[Z_0, U^{-1}\delta U]-
(Z_0 -U^{-1} Z_0 U)[\bar Z_0, U^{-1}\delta U]\right] \\ \nonumber
&=& -16\pi\theta\lambda i \int \delta\sigma\, \sigma {\rm Tr}[Z_0, \bar Z_0]U^{-1}\delta U +
8\pi\theta\lambda i \int \delta\sigma\, \sigma \delta {\rm Tr}\left(\bar Z_0 U Z_0 U^{-1} - Z_0 U \bar Z_0 U^{-1} \right)
\end{eqnarray}
The last term integrates to ${\rm Tr}\left(\bar Z_0 U Z_0 U^{-1} - Z_0 U \bar Z_0 U^{-1} \right)$ at $\tau =0$ and $\tau =1$.
Since $U\equiv {\bf 1}$ at these points, this term should give zero. Therefore,
\begin{eqnarray}
\int \Omega &=& -16\pi\theta\lambda i \int \delta\sigma\, \sigma {\rm Tr}[Z_0, \bar Z_0]U^{-1}\delta U \\ \nonumber
& = & - 8\pi\theta\lambda i \int {\rm Tr}\left[([Z_0, \bar Z_0]+\frac{1}{2\theta})U^{-1}\delta U \right] +
4\pi\lambda i \int_0^1 d\tau {\rm Tr}U^{-1}d_{\tau}U
\end{eqnarray}
and we see that if the Gauss law constraint (\ref{Aeq}) is satisfied, then the first term disappears and we are left with
\begin{equation}
\int \Omega = 4\pi\lambda i \int_0^1 d\tau {\rm Tr}U^{-1}d_{\tau}U.
\end{equation}
only. As it was shown previously \cite{Nair}, $\Pi_1({\cal G}) = {\mathbb Z}$ and
\begin{equation}
Q[U] = \frac{i}{2\pi} \int_0^1 d\tau {\rm Tr}U^{-1}d_{\tau}U
\end{equation}
is an integer equal to the winding number of the class in $\Pi_1({\cal G})$ to which $U(\tau)$ belongs. Also from the general
principles of geometric quantization we know that the reduced phase space can be quantized only if the integral of
$\Omega/{2\pi}$ over any closed noncontractible surface is
an integer. Therefore we can write the following quantization condition
\begin{equation}
4\pi\lambda Q[U] = {\rm integer}
\end{equation}
which can  be satisfied for arbitrary $Q[U]\in {\mathbb Z}$ only if the level number is quantized as
\begin{equation}\label{quantcond}
4\pi\lambda = k \ \ \ \ \ k=0, \pm 1, \ldots
\end{equation}
This is exactly the same quantization condition as was found in \cite{Nair} using Lagrangian approach to the
noncommutative Chern-Simons theory.
\section{Schr\"odinger representation}
After the preliminary analysis of the classical phase space in the
previous section, we would like now to explicitly quantize our
theory and to show how does the Chern-Simons coefficient quantization condition
(\ref{quantcond}) appear as a requirement of consistency in
implementing the Gauss law (\ref{Aeq}) on physical states.

Canonical quantization of the Poisson structure (\ref{poisson}) leads to the following quantum commutation relations
\begin{equation}\label{quantcomm}
\ [Z_{ij}, \bar Z_{kl}] = -\frac{1}{8\pi\theta\lambda}\,\delta_{il}\delta_{jk}.
\end{equation}
In order to construct a  unitary representation of this canonical algebra, we have to choose polarization on the
phase space of the theory. In this paper we use a holomorphic polarization condition. The wave functionals $\Psi[Z]$
are functionals of $Z$ only; $Z$ is represented trivially as multiplication by $Z$, while $\bar Z$ acts as a functional
derivative with respect to $Z$
\begin{equation}
\bar Z_{ij} \Psi[Z] = \frac{1}{8\pi\theta\lambda} \frac{\delta}{\delta Z_{ij}} \Psi[Z].
\end{equation}
In this representation generator (\ref{gener}) of infinitesimal gauge transformations becomes
\begin{equation}\label{Schrodgen}
G(\phi) = i [\phi, Z]_{ij} \frac{\delta}{\delta Z_{ij}} + 4\pi\lambda i {\rm Tr}\,\phi.
\end{equation}
It is easy to verify that the algebra of these generators closes
\begin{equation}\label{galgebra}
[G(\phi), G(\rho)] = -i G([\phi, \rho])
\end{equation}
provided that we choose $\phi$ and $\rho$ to satisfy
(\ref{gaugecondition}). Closure of this algebra means that
there is no apparent obstruction to demanding that the Gauss law
constraint (\ref{Aeq}) be met by requiring that $G(\phi)$
annihilates physical states
\begin{equation}\label{physwf}
G(\phi) \Psi[Z] = 0.
\end{equation}
However, as it is known from the ordinary commutative Chern-Simons
theory \cite{Bos, Dunne} such condition does not necessarily mean
gauge-invariance of the physical wave-functionals. In fact, in the
commutative case consistent implementation of (\ref{physwf})
requires that the action of the gauge group on states is realized
with a $1$-cocycle which leads to multivalued wave-functionals
unless the level number $\lambda$ is quantized. We want to show now
that similar arguments apply in the noncommutative case as well.

For an arbitrary gauge transformation $g = e^\phi$ its realization on states $\Psi[Z]$ is given by the unitary operator
\begin{equation}
U(g) = e^{-i G(\phi)}.
\end{equation}
As it follows from the definition (\ref{Schrodgen}), we can split $G(\phi)$ as
\begin{equation}
G(\phi) = G_Z(\phi) + 4\pi\lambda i {\rm Tr}\phi,
\end{equation}
where
\begin{equation}
G_Z(\phi) = i [\phi, Z]_{ij} \frac{\delta}{\delta Z_{ij}}
\end{equation}
is the generator of infinitesimal gauge transformations on $Z$.  Therefore,
\begin{eqnarray}\label{gtrans}
\Psi[Z]\ \longrightarrow\ U(g)\Psi[Z]\ &=&\ e^{-i G(\phi)} e^{i G_Z(\phi)} \Psi[Z^g]\\ \nonumber
Z^g &=& g Z g^{-1}.
\end{eqnarray}
The prefactor $e^{-i G(\phi)} e^{i G_Z(\phi)}$ can easily be evaluated since
$[G_Z(\phi), 4\pi\lambda i\, {\rm Tr}\,\phi] = 0$ and the result is just $e^{4\pi\lambda\, {\rm Tr}\,\phi}$.
The Gauss law constraint (\ref{physwf}) requires that physical states $\Psi_{phys}[Z]$ be left unchanged by the action of $U(g)$, since
the generator $G(\phi)$ annihilates them (\ref{physwf})
\begin{equation}
U(g) \Psi_{phys}[Z] = \Psi_{phys}[Z].
\end{equation}
Therefore, in the noncommutative Chern-Simons theory, functionals
describing physical states are not gauge invariant; rather,
according to (\ref{gtrans}), they satisfy
\begin{equation}\label{gconst}
\Psi_{phys}[Z^g] = e^{-4\pi\lambda {\rm Tr}\phi} \Psi_{phys}[Z].
\end{equation}
However, the above expression can not be met with single-valued
functionals unless $\lambda$ is quantized. To see this it is useful to rewrite (\ref{gconst}) in the following equivalent
way 
\begin{equation}\label{gconstt}
\Psi_{phys}[Z^g] = ({\rm det}\, g)^{-4\pi\lambda}\ \Psi_{phys}[Z].
\end{equation}
As was argued in the previous section, the valid gauge
transformations are given by essentially finite (although they can be very large) unitary matrices $g$. For such matrices 
${\rm det}\, g$ is well-defined (it is basically a complex number with unit modulus). However, $({\rm det}\, g)^{-4\pi\lambda}$
is multivalued unless the exponent $-4\pi\lambda$ is an integer. Therefore, for (\ref{gconst}) to make sense, $\lambda$
has to be quantized in units of $\frac{1}{4\pi}$
\begin{equation}
4\pi\, \lambda = k, \ \ \ \ \ \ \ \ \ k = 0, \pm 1, \ldots
\end{equation}
so again we obtain the same quantization condition as (\ref{quantcond}).

Finally, to have a well-defined quantum theory we need to define
the inner product on the Hilbert space of physical states. The inner
product of two wave functionals is given by
\begin{equation}\label{innerprod}
\langle\Phi|\Psi\rangle = \int [dZ, d\bar Z]\, e^{-8\pi\theta\lambda \,{\rm Tr}\,\bar Z Z}\Phi^{\star} [\bar Z]\Psi [Z]
\end{equation}
where the exponential prefactor is just the K\"ahler potential
(\ref{Kahler}), as is standard in holomorphic quantization. This
prefactor ensures that $\bar Z$ is the hermitian conjugate of $Z$,
the quantum version of the classical relation $\bar Z =
{(Z)}^\dagger$. Also it can be easily verified that this inner
product is insensitive to the gauge noninvariance of states
meaning that physical expectation values do not depend on the
gauge choice as they should.
\section{Physical states}
Given that we know the gauge transformation properties of the physical states, we now want to explicitly construct
functionals that obey (\ref{gconst}). But before we proceed to the details of this construction, we would like to briefly
outline our strategy.
 Equation (\ref{gconstt}) 
tells us, that under the gauge transformation $g$ the wave
functionals are multiplied by some power of ${\rm det}\, g$.
Therefore, if $h$ is some noncommutative matrix field
parametrizing covariant derivative $Z$ and transforming as
\begin{equation}\label{htrans}
h \longrightarrow h^g = g h,
\end{equation}
then $({\rm det}\, h)^{-4\pi\lambda}$ transforms exactly as (\ref{gconstt}), and the most general functional with the correct
transformation properties is given by
\begin{equation}\label{vwfdet}
\Psi_{\rm phys}[Z] = ({\rm det}\, h)^{-4\pi\lambda}\, \psi[Z].
\end{equation}
Here $\psi[Z]$ is an arbitrary gauge-invariant functional of $Z$ only. In the case of pure Chern-Simons theory on the
noncommutative plane the only such functional is $\psi[Z] \sim 1$ so the physical Hilbert space of the theory is
one-dimensional with the only vacuum state given (up to normalization) by
\begin{equation}\label{vacstate}
\Psi_{VAC}[Z] = ({\rm det}\, h)^{-4\pi\lambda}.
\end{equation}

To make these heuristic arguments precise we need, first of all,
parametrization of the covariant derivative $Z$ in terms of the
matrix field $h$ obeying (\ref{htrans}) and then, we need to give
an exact meaning to ${\rm det}\, h$ since, unlike the gauge uransformations $g$, $h$ is an infinite
matrix and it is not clear {\textit {a priori}} what ${\rm det}\,
h$ is in this case and whether the usual property of determinants
${\rm det}\, h^g = ({\rm det}\, h)({\rm det}\, g$), which is crucial in deriving (\ref{vacstate}),  still holds in
the noncommutative case.

In the commutative field theory in two space dimensions, the following parametrization of the gauge potential $A_z$ is
frequently used
\begin{equation}\label{commutpar}
A_z = -\partial_{z} h h^{-1}.
\end{equation}
The reason why such parametrization is possible is that in two
dimensions operator $\partial_z$ is invertible and, for any field
configuration $A_z$ we can invert (\ref{commutpar}) and find
corresponding $h$ at least perturbatively as a series in powers of
$A_z$. It is also easy to verify, that under the gauge
transformations $h$ transforms as in (\ref{htrans}).

With these ideas in mind, we introduce the following parametrization of the noncommutative covariant derivative
\begin{equation}\label{mainpar}
Z = \frac{i}{2\theta}\, h \bar z h^{-1}.
\end{equation}
One can use the relationship (\ref{DArelation}) between covariant derivative $Z$ and
noncommutative gauge potential $A_z$ as well as the definition of the noncommutative derivatives
(\ref{ncder}) to see that (\ref{mainpar}) is the noncommutative
analogue of (\ref{commutpar}). It is shown in the Appendix that
(\ref{mainpar}) gives the valid parametrization in the sense that
it can be perturbatively solved for $h$. 

Now we have to clarify the meaning of ${\rm det}\, h$ in (\ref{vacstate}) and to do that we start with the well-known
expression
\begin{equation}
{\rm det}\, h = e^{{\rm Tr\, log}\, h}
\end{equation}
which we use to put ${\rm det}\, h$ into an exponential form. Then ${\rm Tr\, log}\, h$ can be written as
\begin{equation}\label{intform}
{\rm Tr\, log}\, h = \int_0^1 d\tau\, {\rm Tr}\,({\tilde h}^{-1} \partial_{\tau} \tilde h)
\end{equation}
where $\tilde h (\tau)$ is the smooth extension of $h$ onto the line segment $\tau \in [0, 1]$ with the following values
at the boundary
\begin{eqnarray}
\tilde h (0) & \equiv &{\bf 1}\\\nonumber
\tilde h (1) & \equiv & h.
\end{eqnarray}
One can use a power series expansion of $\tilde h (\tau) = e^{\chi (\tau)}$ to verify (\ref{intform}). Under the gauge
transformation $U = e^{\phi}$ the  value of $\chi (\tau)$ on the $\tau = 1$ boundary transforms as
$\chi^U (1) = \chi (1)+ \phi + [\phi, \chi (1)] +\ldots$ and if $\phi$ goes sufficiently fast to zero at infinity
(as required by (\ref{gaugecondition})) then under the trace all the commutator terms vanish and we get
\begin{equation}
e^{{\rm Tr\, log}\, h^U} = e^{{\rm Tr\, log}\,h + {\rm Tr}\, \phi}.
\end{equation}
This is exactly the transformation we need for the physical states.
As a result we see that (\ref{intform}) has all the required transformation properties and therefore can be used to write
the vacuum state as
\begin{equation}
\Psi_{VAC}[Z] = e^{-4\pi\lambda \int_0^1 d\tau {\rm Tr}\,({\tilde h}^{-1} \partial_{\tau} \tilde h)}.
\end{equation}

Although this expression can already be used as the definition of  $\Psi_{VAC}[Z]$, we would like to explore it a bit further.
In particular, we want to show how it is related to the noncommutative WZW action
\begin{equation}
S_{NCWZW} = S_{KIN} + S_{NCWZ} = 2\pi\theta {\rm Tr}\,(h^{-1}\partial_z h) (h^{-1}\partial_{\bar z} h) +
 2\pi\theta \int d\tau {\rm Tr}\, \left( h^{-1}\partial_{\tau} h \left[ h^{-1}\partial_z h, h^{-1}\partial_{\bar z} h \right]
\right).
\end{equation}
Using the definition of derivatives on the noncommutative plane (\ref{ncder}) and formally expanding the commutators we can transform
the Wess-Zumino term as
\begin{eqnarray}\label{eqWZ}
S_{NCWZ} &=& 2\pi\theta \int d\tau {\rm Tr} \left( h^{-1}\partial_{\tau} h \left[ h^{-1}\partial_z h, h^{-1}\partial_{\bar z} h \right]
\right)\\ \nonumber
&=& \frac{\pi}{2\theta}\int d\tau {\rm Tr} \left( h^{-1}\partial_{\tau} h \left[ h^{-1}[z, h], h^{-1}[\bar z, h] \right] \right)\\\nonumber
&=& 2\pi \int d\tau {\rm Tr}\left( h^{-1}\partial_{\tau} h \right) -
\frac{\pi}{2\theta}\int d\tau {\rm Tr}\, h^{-1}\partial_{\tau} h \left([h^{-1}z h, \bar z] + [z, h^{-1}\bar z h]\right)\\\nonumber
&=& 2\pi \int d\tau {\rm Tr}\left( h^{-1}\partial_{\tau} h \right) +
\frac{\pi}{2\theta}\int d\tau \partial_{\tau}{\rm Tr}\,(h^{-1}z h \bar z - h^{-1}\bar z h z )\\\nonumber
&=& 2\pi \int d\tau {\rm Tr}\left( h^{-1}\partial_{\tau} h \right)
+ \frac{\pi}{2\theta}{\rm Tr}(h^{-1}z h \bar z - z\bar z)- \frac{1}{4\theta^2} {\rm Tr} (h z h^{-1} \bar z - z\bar z).
\end{eqnarray}
Similarly the kinetic term becomes
\begin{eqnarray}\label{eqKIN}
S_{KIN} &=& 2\pi\theta {\rm Tr}(h^{-1}\partial_z h) (h^{-1}\partial_{\bar z} h)\\\nonumber
&=& -\frac{\pi}{2\theta} {\rm Tr} (h^{-1}z h - z)(h^{-1}\bar z h - \bar z)\\\nonumber
&=& \frac{\pi}{2\theta} {\rm Tr} (h^{-1}z h \bar z - z\bar z) + \frac{\pi}{2\theta} {\rm Tr} (h z h^{-1} \bar z - z\bar z).
\end{eqnarray}
Note that we did not use the cyclicity of trace while performing these transformations. Also, although each of the 
expressions like ${\rm Tr}\, z\bar z$ or ${\rm Tr}\, h^{-1}z h \bar z$ is divergent, their difference, as it appears in the
last lines of (\ref{eqWZ}) and (\ref{eqKIN}), is, in fact, a well-defined finite quantity. These two simple observations in 
certain sense validate the formal manipulations that we have done. 
  
Now, if we put the two terms together, we get the following identity
\begin{equation}
2\pi \int d\tau {\rm Tr} h^{-1}\partial_{\tau} h =  S_{NCWZW} + 2\pi i{\rm Tr}(zA)
\end{equation}
which can be used to rewrite the vacuum state as
\begin{equation}\label{VWFfinal}
\Psi_{VAC}[Z] = e^{-2\lambda S_{NCWZW} -4\pi\lambda i {\rm Tr}(zA)}.
\end{equation}
This expression is much more convenient for the analysis of the transformation properties of $\Psi_{VAC}[Z]$ since now we can use
the well-known Polyakov-Wiegmann identity \cite{Polyakov} (which still holds in the noncommutative theory)
\begin{equation}
S_{NCWZW}(g h) = S_{NCWZW}(g) + S_{NCWZW}(h) + 2\int {\rm Tr}\,(g^{-1}\bar\partial g \partial h h^{-1})
\end{equation}
to prove that (\ref{VWFfinal}) transforms properly under the unitary gauge transformations.
Also it is easy to see that our parametrization of $Z$ in terms of $h$ is somewhat ambiguous. Really, if $h$ is some solution
of (\ref{mainpar}), then $h f(\bar z)$, where $f(\bar z)$ is
some antiholomorphic function, gives an equivalent solution of
that equation and can be used to define the vacuum state
of the theory as well. $\Psi_{VAC}[Z]$ should
certainly be invariant with respect to such ambiguity in the
choice of parametrization and again we can use the Polyakov-Wiegmann identity to see that this is indeed the case.

Finally the above expression for the ground state looks very similar to the well-known vacuum wave functional of the commutative
Chern-Simons theory \cite{Bos}
\begin{equation}\label{VWFcommut}
\Psi_{VAC_{commut}}[A] = e^{-2\lambda S_{WZW}(h)}
\end{equation}
except for the last term in the exponential of (\ref{VWFfinal}). The reason for appearance of such term can be easily
tracked down to our choice of the covariant derivative operators $Z, \bar Z$ as the fundamental set of variables of the
theory. Really, the change in the phase space parametrization from $Z, \bar Z$ to $A, \bar A$
is essentially the canonical shift of variables according to (\ref{DArelation}),. Upon such transformation, the path-integral measure
in the inner product (\ref{innerprod}) becomes
\begin{equation}
\int [ dZ ] [ d\bar Z ]\, e^{-8\pi\theta\lambda\, {\rm Tr}\, (\bar Z Z)} =
\int [ dA ] [ d\bar A ]\, e^{-8\pi\theta\lambda\, {\rm Tr}\, (\bar A A) + 4\pi\lambda i {\rm Tr}\, (z A)
-4\pi\lambda i {\rm Tr}\, (\bar z\bar A) -\frac{2\lambda}{\theta} {\rm Tr}\, (z\bar z)}
\end{equation}
and we see, that $e^{4\pi\lambda i\, {\rm Tr}\, (z A)}$ can be absorbed into $\Psi[Z]$ thus cancelling the extra
term in the wave function
($e^{-4\pi\lambda i\, {\rm Tr}\, (\bar A\bar z)}$ correspondingly is absorbed by $\bar\Phi[\bar Z]$ in (\ref{VWFfinal})).
Therefore, the canonically transformed wave functional of variable $A$ is
\begin{equation}
\Psi_{VAC}[A] = e^{-2\lambda S_{NCWZW}(h)}.
\end{equation}
In the commutative limit $\theta\to 0$ the noncommutative WZW action $S_{NCWZW}(h)$ goes to the commutative one
$S_{WZW}(h)$ and we trivially recover the ground state of the commutative Chern-Simons theory (\ref{VWFcommut}). Similarly,
the only remaining term \footnote{The other term $-\frac{2\lambda}{\theta} {\rm Tr}\, (z\bar z)$, being independent of $A$,
can be absorbed into the wavefunction normalization constant.} $-8\pi\theta\lambda\, {\rm Tr}\, (\bar A A)$ in the Hilbert
space measure (\ref{innerprod}) becomes just
$-4\lambda \int d^2 z \, {\rm Tr}\, (\bar A A)$ in the limit of vanishing noncommutativity and again we obtain the standard expression for the inner product of states in
the commutative case
\begin{equation}
\langle\Phi |\Psi\rangle = \int [ dA ] [ d\bar A ]\, e^{-4\lambda\, {\rm Tr}\, (\bar A A)}
 \Phi^{\star}[\bar A] \Psi [A].
\end{equation}
\section{Summary and conclusions}
In this paper the hamiltonian analysis of the pure $U(N)$
Chern-Simons theory on the noncommutative plane was performed. It
was found that quantization of the level number in the canonical
formalism is a consequence of existence of the closed
noncontractible surfaces in the reduced phase space of the theory.
The quantization condition (\ref{quantcond}) is exactly the same
as was previously obtained in \cite{Nair} using Lagrangian
approach. Also like its commutative counterpart, pure
noncommutative CS theory turns out to be exactly solvable. We use
the techniques of holomorphic quantization to construct an
explicit representation of the quantum commutator algebra
(\ref{quantcomm}). Furthermore, it is shown that the Gauss law
constraint (\ref{Aeq}), which selects the subspace of physical
states of our theory, can be solved exactly.  The physical Hilbert
space for our choice of flat space geometry is found to be
one-dimensional and we give an explicit expression for the only
physical state of the theory.

Although pure Chern-Simons theory appears to be trivial, it is well-known that in the commutative case it leads to
highly nontrivial results when coupled to external sources. Therefore, it appears to be  interesting to include  external
charges into the noncommutative theory as well. In particular, one might address the question of how does the presence of
such charges affect the quantum holonomy of physical states. This is currently under investigation.

\acknowledgements{The author would like to thank Prof.~V.P.~Nair for suggesting this problem, numerous discussions
and advise during the work on this project as well as Prof.~A.~P.~Polychronackos for clarifying some critical issues.}
\appendix
\section{}
In this Appendix we would like to show that parametrization
(\ref{mainpar}) of the covariant derivative operator $Z$ in terms
of an infinite matrix $h$ is well-defined in the sense that for any given
matrix $Z$ it is possible, at least perturbatively, to find
corresponding matrix $h$. It is useful to rewrite (\ref{mainpar}) in
an equivalent form
\begin{equation}\label{Aeeq}
A = -\frac{-i}{2\theta}\,[\bar z, h]\, h^{-1},
\end{equation}
where $A$ is the noncommutative gauge potential as defined in (\ref{DArelation}). To be able to solve this equation we have to prove
that operator $\frac{-i}{2\theta}[\bar z, \ldots]$ is invertible, i.e. that there exists a map (we
call it $D(\ldots)$)
\begin{equation}\label{map}
B \stackrel{D}{\longrightarrow} D(B)
\end{equation}
which associates to any given noncommutative function $B$ another function $D(B)$ such that
\begin{equation}\label{meq}
-\frac{i}{2\theta}[\bar z, D(B)] = B.
\end{equation}
This map is the noncommutative analogue of $\int dw\, G(z,w)\ldots$ with $G(z,w)$ being the Green's function of
the ordinary commutative derivative operator $\partial_z$. In terms of $D(\ldots)$ we can write then the solution of (\ref{Aeeq}) as 
\begin{equation}\label{qwe}
h = 1 + D(Ah) = 1 + D(A) + D(A D(A)) + \ldots.
\end{equation}
However, the validity of this expression crucially depends on the existence of map $D(\ldots)$, so we give now the proof that
$D(\ldots)$ is indeed a well-defined operation on the noncommutative plane.
  
Since $B$ and $D(B)$ are both infinite-dimensional matrices, we
can represent them in the oscillator basis as
\begin{eqnarray}
B &=& \sum_{i,j = 0}^{\infty} B_{ij}|i\rangle\langle j|\\\nonumber
D(B) &=& i\sqrt{2\theta} \sum_{i,j = 0}^{\infty} C_{ij}|i\rangle\langle j|
\end{eqnarray}
With this expansion eq.(\ref{meq}) now gives the following set of
recursion relations for matrix elements of $D(B)$
\begin{equation}\\\label{recurcion}
C_{i-1\, j}\sqrt{i} - C_{i\, j+1}\sqrt{j+1} = B_{ij} \ \ \ \ \ \ i,j=0,1,2,\ldots
\end{equation}
From these we find
\begin{eqnarray}
B_{00} &=& -C_{01}\\\nonumber
B_{01} &=& -C_{02}\sqrt{2}\\\nonumber
&\ldots\ldots\\\nonumber
B_{0l} &=& - C_{0 l+1}\sqrt{l+1},
\end{eqnarray}
which means that we can immediately find all $C_{0l}$ coefficients with $l\geq 1$. Next we consider the following set of equations
\begin{eqnarray}
B_{11} &=& C_{01} - C_{12}\sqrt{2}\\\nonumber
B_{12} &=& C_{02} - C_{13}\sqrt{3}\\\nonumber
&\ldots\ldots\\\nonumber
B_{1l} &=& C_{0l} - C_{1 l+1}\sqrt{l+1}
\end{eqnarray}
and obtain $C_{1l}$, $l\geq 2$. Now we can proceed iteratively and see that given that we have already found $C_{i-1l}, l\geq i$
for some $i$, we can always find $C_{il}, l\geq i+1$ from
\begin{eqnarray}
B_{ii} &=& C_{i-1 i}\sqrt{i} - C_{i i+1}\sqrt{i+1}\\\nonumber
&\ldots\ldots\\\nonumber
B_{il} &=& C_{i-1 l}\sqrt{i} - C_{i l+1}\sqrt{l+1}.
\end{eqnarray}
Therefore, all the matrix elements $C_{ij}$ of $D(B)$ with $i<j$
can be uniquely determined from the above equations.

For those $C_{ij}$ with $i\geq j$ we may consider the following set of equations
\begin{eqnarray}
B_{10} &=& C_{00} - C_{11}\\\nonumber
B_{21} &=& C_{11}\sqrt{2} - C_{22}\sqrt{2}\\\nonumber
&\ldots\ldots\\\nonumber
B_{i+1 i} &=& C_{ii}\sqrt{i+1} - C_{i+1 i+1}\sqrt{i+1}.
\end{eqnarray}
From these equations we can find all diagonal matrix elements $C_{ii},\, i\geq0$ provided that we fix arbitrarily the value
of $C_{00}$. Easy to see that this freedom in choosing $C_{00}$ translates into the following ambiguity of $D(B)$
\begin{equation}
C_{00} \sum_{0}^{\infty} |i\rangle\langle i| = C_{00} {\bf 1}
\end{equation}
so we can add an arbitrary constant function to $D(B)$. Similarly, from
\begin{eqnarray}
B_{20} &=& C_{10}\sqrt{2} - C_{21}\\\nonumber
B_{31} &=& C_{21}\sqrt{3} - C_{32}\sqrt{2}\\\nonumber
&\ldots\ldots\\\nonumber
B_{i+2 i} &=& C_{i+1 i}\sqrt{i+2} - C_{i+2 i+1}\sqrt{i+1}
\end{eqnarray}
we can find all $C_{i+1 i},\, i\geq 0$, however, solution is not unique again; we can add
\begin{equation}
C_{10} \sum_{0}^{\infty} \sqrt{i+1}|i+1\rangle\langle i| = C_{10} \bar z
\end{equation}
to $D(B)$. In exactly the same way one can show that all the
remaining matrix elements $C_{ij}$ with $i\geq j$ can be found
from (\ref{recurcion}) and this completes the proof of existence
of the map (\ref{map}). This map, however, is not unique; D(B) is
defined up to
\begin{equation}
C_{00}{\bf 1} + C_{10} \bar z + C_{20} {\bar z}^2 + C_{30} {\bar z}^3 +\ldots
\end{equation}
with arbitrary coefficients ${C_{00}, C_{10}, \ldots}$, {\textit
i.e.} we can add any noncommutative antiholomorphic function $f(\bar z)$ to
$D(B)$ and still satisfy (\ref{meq}).

 One can also see that solution (\ref{qwe}) of equation (\ref{Aeeq}) is not unique as well. Really, because of the ambiguity in the 
definition of $D(\ldots)$ we can write an alternative solution of (\ref{Aeeq}) as
\begin{equation}
h = f(\bar z) + D(A h) = f(\bar z) + D\left(A f(\bar z)\right) + D\left(A D\left(A f(\bar z)\right)\right)+\ldots 
\end{equation}
However, this can also be written as
\begin{equation}
h =\left[1 + D(A f(\bar z))f^{-1}(\bar z) +\ldots \right] f(\bar z) = \left[1 + D^f(A) + D^f(A D^f(A)) +\ldots \right] f^{-1}(\bar z) 
\end{equation}
where $D^f(\ldots) = D(\ldots f(\bar z)) f^{-1}(\bar z)$ satisfies (\ref{meq}) and expression in brackets
\begin{equation}
h^f = 1 + D^f(A) + D^f(A D^f(A)) +\ldots 
\end{equation}
is the solution of (\ref{Aeeq}) as well. Therefore, we see that if $h$ is some solution of (\ref{Aeeq}) then
\begin{equation}
h^f = h f(\bar z)
\end{equation} 
is another solution of that equation. This means that our parametrization of the covariant derivative $Z$ in terms of the matrix
field $h$ is defined up to right multiplication by an arbitrary antiholomorphic function only. In fact, this can be
seen directly from (\ref{mainpar}) since any such function obviously commutes with the antiholomorphic coordinate operator~
$\bar z$. 



\end{document}